\documentclass[fleqn,10pt]{article}
\usepackage[utf8]{inputenc}

\usepackage{cite}
\usepackage{amsmath,amssymb,amsfonts}
\usepackage{graphicx}
\usepackage{textcomp}
\usepackage{xcolor}
\usepackage{subcaption}
\usepackage{gensymb}
\usepackage{float}
\usepackage{algorithm,algcompatible,algpseudocode}
\usepackage{array}
\usepackage{hyperref}
\usepackage{forest}
\usepackage{booktabs}
\usepackage{authblk}

\def\BibTeX{{\rm B\kern-.05em{\sc i\kern-.025em b}\kern-.08em
    T\kern-.1667em\lower.7ex\hbox{E}\kern-.125emX}}

\begin{document}

\title{An Indoor Radio Mapping Dataset Combining 3D Point Clouds and RSSI
}

\author[1,2]{Ljupcho Milosheski}
\author[3]{Kuon Akiyama}
\author[1]{Bla\v{z} Bertalani\v{c}}
\author[1]{Jernej Hribar}
\author[3]{Ryoichi Shinkuma}
\affil[1]{Jožef Stefan Institute, Department of Communication Systems, Ljubljana, 1000, Slovenia}
\affil[2]{International Postgraduate School Jožef Stefan, Information and Communication Technologies, Ljubljana, 1000, Slovenia}
\affil[3]{Faculty of Engineering, Shibaura Institute of Technology, 3 Chome-7-5 Toyosu, Koto City, Japan}
\affil[*]{Corresponding author: ljupcho.milosheski@ijs.si}

\maketitle

\begin{abstract}

The growing number of smart devices supporting bandwidth-intensive and latency-sensitive applications, such as real-time video analytics, smart sensing, Extended Reality (XR), etc., necessitates reliable indoor wireless connectivity. {In such environments}, accurate Radio Environment Maps (REMs) enable adaptive wireless network planning and optimization of Access Point (AP) placement. {However, generating realistic REMs remains difficult due to the variability of indoor environments and the limitations of existing modeling approaches, which often rely on simplified layouts or synthetic data. These challenges are further amplified by the adoption of next-generation Wi-Fi standards, operating at higher frequencies with limited range and wall penetration. To support progress in this area, we collected a dataset that combines high-resolution 3D LiDAR scans with Wi-Fi RSSI measurements across 20 setups in a multi-room indoor environment.} It includes two measurement scenarios, one with and one without human presence, {enabling development and validation of REM estimation models that incorporate physical geometry and environmental dynamics.} The described dataset supports research in data-driven wireless modeling and the {development of} high-capacity indoor {communication} networks.

\end{abstract}

\textbf{Keywords:} indoor radio mapping, LiDAR, RSSI, wireless, dataset

%  Click the title above to edit the author information and abstract
%
\thispagestyle{empty}

\section*{Background \& Summary} % / Introduction}

Smart devices such as cameras, doorbells and voice assistants like Alexa have become an integral part of our daily lives, as they make everyday tasks much easier and increase productivity. To provide smart services, these devices require a continuous internet connection and their bandwidth requirements are expected to increase with the integration of advanced features such as high-definition video streaming and real-time Artificial Intelligence (AI) inference tasks. To fulfil these requirements in an indoor environments, wireless technologies such as Wi-Fi are particularly well suited due to their high data rates, cost efficiency and adaptability. However, optimising Wi-Fi performance in complex indoor environments remains a challenge due to factors such as signal attenuation from walls, interference from household appliances and dynamic physical obstructions. This becomes even more urgent with the rise of Extended Reality (XR) technologies, e.g., Virtual Reality (VR), Mixed Reality (MR), Augmented Reality (AR)~\cite{akyildiz2022wireless}, etc., which impose far more stringent connectivity requirements such as ultra-low latency and extremely high throughput~\cite{rohde2023xr}. Their demands often exceed those of conventional smart devices and expose the limitations of existing Wi-Fi deployments, particularly when Access Points (APs) are suboptimally placed. The challenge is further amplified with the deployment of newer standards like Wi-Fi 7 (IEEE 802.11be) and the upcoming Wi-Fi 8 (IEEE 802.11bn), whose use of higher frequencies results in reduced wall penetration and limited range.

Consequently, the development of a comprehensive dataset documenting Wi-Fi propagation characteristics in indoor environments, for example homes with multiple rooms or office spaces, is critical. Such datasets will enable the construction of detailed indoor Radio Environment Maps (REMs), support the development of predictive channel models and AI-driven network optimization tools, and inform the strategic placement of APs to sustain high-performance connectivity. This is particularly important for the design of XR-ready environments where immersive and seamless experiences must be reliably maintained in real time.

Various approaches for REMs have been proposed in the literature, which are crucial for understanding spatial signal distribution and optimising wireless deployment. The classical techniques can be broadly categorised into direct methods, which spatially interpolate signal measurements\cite{pesko2014radio}, and indirect methods, which are based on known or estimated transmission parameters. Among the most established are ray-tracing-based models implemented in tools such as Wireless InSite \cite{wireless_insite} and NVIDIA’s Sionna \cite{sionna}, which simulate electromagnetic propagation using geometric and physical scene representations. These methods are particularly useful for generating large synthetic datasets by varying the placement of APs and environmental configurations, providing a valuable basis for training supervised learning models. In parallel, Machine Learning (ML) based approaches have gained significant attention due to their ability to learn complex propagation patterns from data. Classical ML methods such as Support Vector Machines (SVMs) and gradient boosting algorithms such as XGBoost \cite{rufaida2020construction} have been used for REM construction, while more recently deep learning techniques such as Generative AI (GAI) \cite{wang2024radiodiff}, Large Language Models (LLMs) \cite{quan2025large} and Graph Neural Networks (GNNs) \cite{bufort2023gnn, li2025radiogat}, have shown promise for learning high-dimensional mappings between environmental inputs and signal properties. In particular, GAI has shown the potential to combine heterogeneous data modalities, such as geometric information and wireless signal features, which is crucial for realistic REM generation. While many of these methods have been evaluated in both outdoor \cite{chen2024act, jaensch2024radio} and indoor \cite{bakirtzis2022deepray, bakirtzis2024rigorous, cisse2024fine} environments, the lack of publicly available, high-resolution datasets remains a significant limitation in development advanced methods for optimising wireless deployments. % Existing datasets often lack the spatial diversity required for modeling complex indoor propagation, especially for high performance applications such as XR. 
Combining Received Signal Strength Indicator (RSSI) measurements with detailed 3D point cloud data offers a way to overcome this gap by embedding the physical geometry directly into the signal prediction. This representation enables learning-based models to account for occlusions, materials and structural features that influence signal behaviour, supporting the development of more accurate, robust and generalisable SEM estimation systems tailored to future wireless connectivity requirements.

REM models development for indoor use cases, as a target purpose of the data published in this paper, is typically solved based on floor plans and material properties \cite{cisse2023irgan}, considering a simpler, top-view 2D scenario. Recently, more detailed 3D data of the environments \cite{bakirtzis2024indoor}, such as furniture and room geometry, are also part of the input, thus providing the advantage of containing fine-grain specifics of the 3D environment. However, the developed indoor models often rely on purely simulated datasets for training and performance validation due to the labor-intensive process of performing actual measurements and the necessity of large datasets for training. While ray-tracing tools could provide highly realistic simulated data in large quantities, thus solving the issue of the data requirement, they cannot fully replicate the complex and dynamic nature of real indoor signal propagation.

In this paper, we describe a dataset that combines high-resolution 3D point clouds of an indoor environment, captured using LiDAR sensors \cite{akiyama2023real}, and corresponding Wi-Fi RSSI measurements, {when using single AP}. The 3D scans can be used to construct geometrically accurate models, compatible with simulation frameworks such as Mitsuba renderer~\cite{nimier2019mitsuba}, supporting hybrid approaches that combine real and synthetic data. {To capture the distinct geometric features that influence signal propagation in indoor environments, we collected measurements in two different indoor areas: an office and a corridor with an adjacent elevator hall. The office space, depicted with orange in Figures 1 and 6a represents open and wide space with office equipment, such as desks and monitors which should serve providing measurements of attenuated signals due to the mentioned obstacles and occupants. The corridor (horizontal blue region on Figure 1) shows long and narrow space, which is prone to reflections due to the small width, and suitable for long line-of-sight data recordings. Finally, the vertical blue region in Figure~1 leading southwards, is another elongated area that contains points which are separated with multiple walls from the points in the office. The whole region of interest spans on $28$m and $30$m in the corresponding vertical and horizontal directions according to Figure 1. The size of the region was made under the assumption that it will be covered by a single AP.} We conducted a measurement campaign to capture the effects of human presence on signal propagation. During this campaign, we collected measurements across two specific scenarios: first in an unoccupied laboratory, and later under realistic conditions with 7–10 individuals engaged in typical activities. The final dataset contains {20} unique {single} AP {setups} and captures the spatial dynamics of signal propagation in a complex indoor environment with and without human presence. By combining physical signal measurements with detailed spatial geometry, we enable analysis of real-world propagation patterns and highlight the impact of environmental variability on the Wi-Fi coverage.

\begin{figure}[!htbp]
        \centering
	\includegraphics[width=0.6\linewidth]{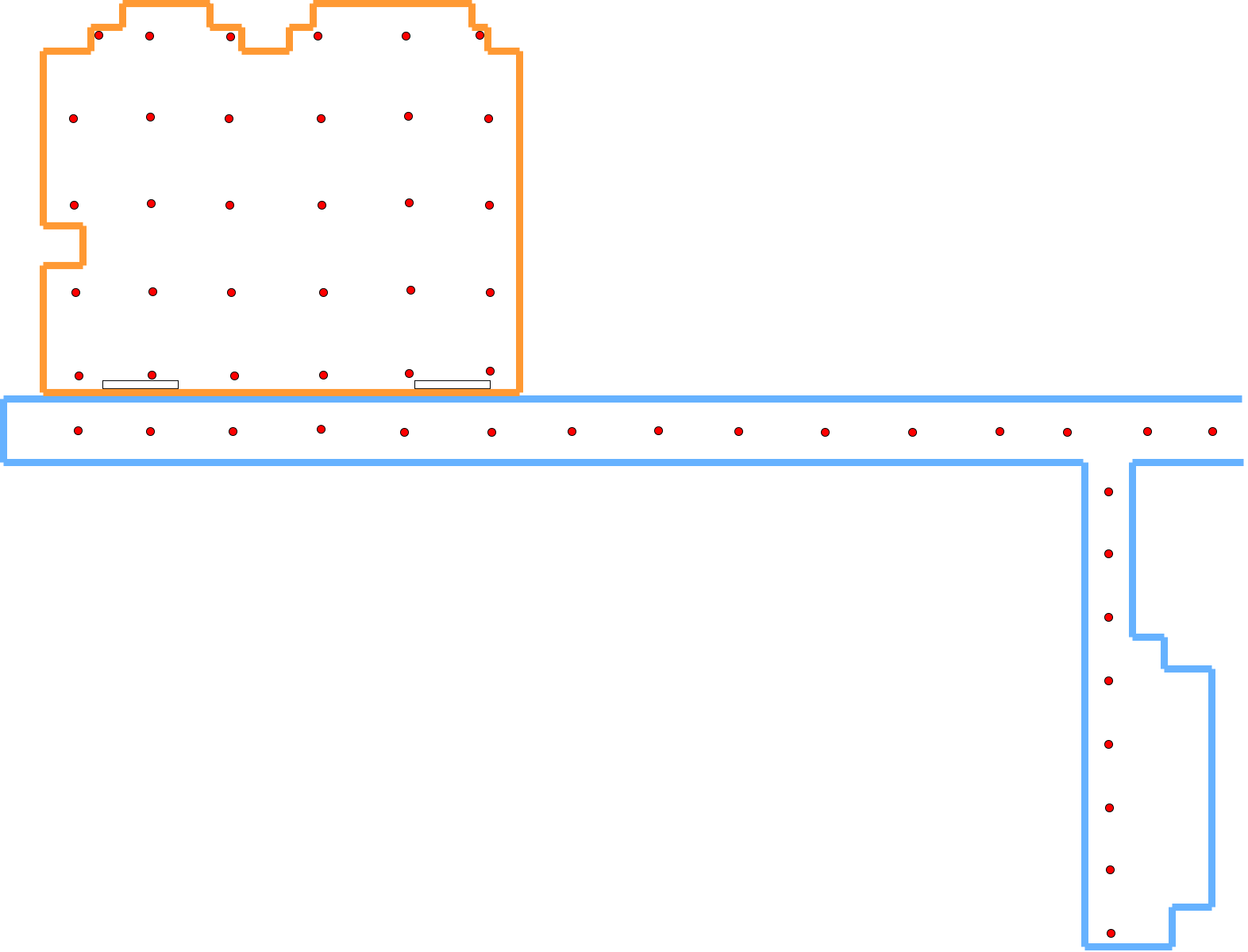}
	\caption{Alignment of the office room and the hallways, including the RSSI measurement locations.}
	\label{fig:indoor_space}
\end{figure}

\section*{Methods}

In this section, we present the procedures and equipment we used for data acquisition.
In the first subsection, we describe how we performed 3D measurements using LiDAR devices; in the second subsection, we explain how we conducted RSSI measurements using a commercial Wi-Fi AC and a user device.

\subsection*{Measurement System for 3D Point Cloud}
\label{sec:measurement_system}

We measured the point clouds of the room using a set of commercially available LiDAR sensors, including VLP-16 units from Velodyne (now Ouster) \cite{vlp16} and Avia units from Livox \cite{avia}. We conducted the measurements in two separate areas: the main room, shown in Figure 2, and the hallway, shown in Figure 3.
In the main room, only the Livox Avia was used. Its extended detection range and relatively high point density made it well-suited for capturing detailed indoor scenes and ensuring accurate reconstruction of furniture and structural elements.
For the hallway, we used both Avia and VLP-16 sensors. The Avia captured long-range details along the corridor with high density, while the VLP-16's 360° horizontal field of view complemented it by covering lateral areas missed by the Avia. Combining both sensors enable for comprehensive point cloud coverage, especially in transitional areas like doorways and corners.
This sensor setup was chosen to leverage each device’s strengths: the Avia for range and density, and the VLP-16 for wide angular coverage.

\begin{figure}[!b]
        \centering
	\includegraphics[width=0.6\linewidth]{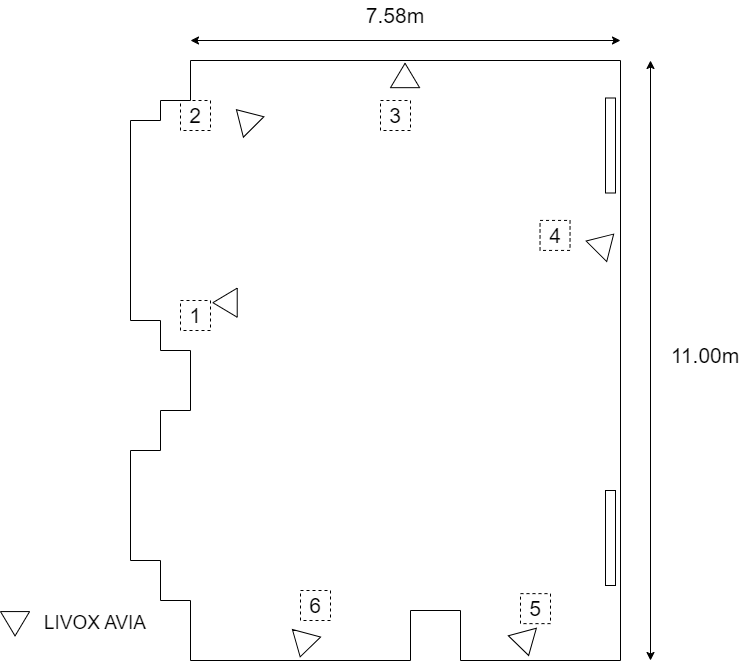}
	\caption{Room with position of LiDAR system.}
	\label{fig:lidar:pos:room}
\end{figure}

In Figure 1, the alignment of the workplace (shown in orange) from Figure 2 and the hallway (shown in blue) from Figure 3 is presented, along with the RSSI measurement locations marked with red dots. In Figure 2, a plan‐view layout of the open‐concept workplace room is presented, annotated with six LiDAR sensor stations around the room’s perimeter. The figure illustrates how the LiDAR sensor array is distributed to provide coverage across the entire workspace. At each labeled station, a triangle marker, representing an Avia LiDAR sensor is shown, while the dashed squares represent the numbering of each station. The floor plan is subdivided into two main zones, a wood‐surfaced area in the upper portion and a tiled area in the lower portion. Stations~1 and~2 are located along the western wall, station~3 is situated at the northern boundary, station~4 appears on the eastern wall, and stations~5 and~6 are along the southern edge.

\begin{figure}[!htbp]
	\centering	\includegraphics[height=0.5\linewidth]{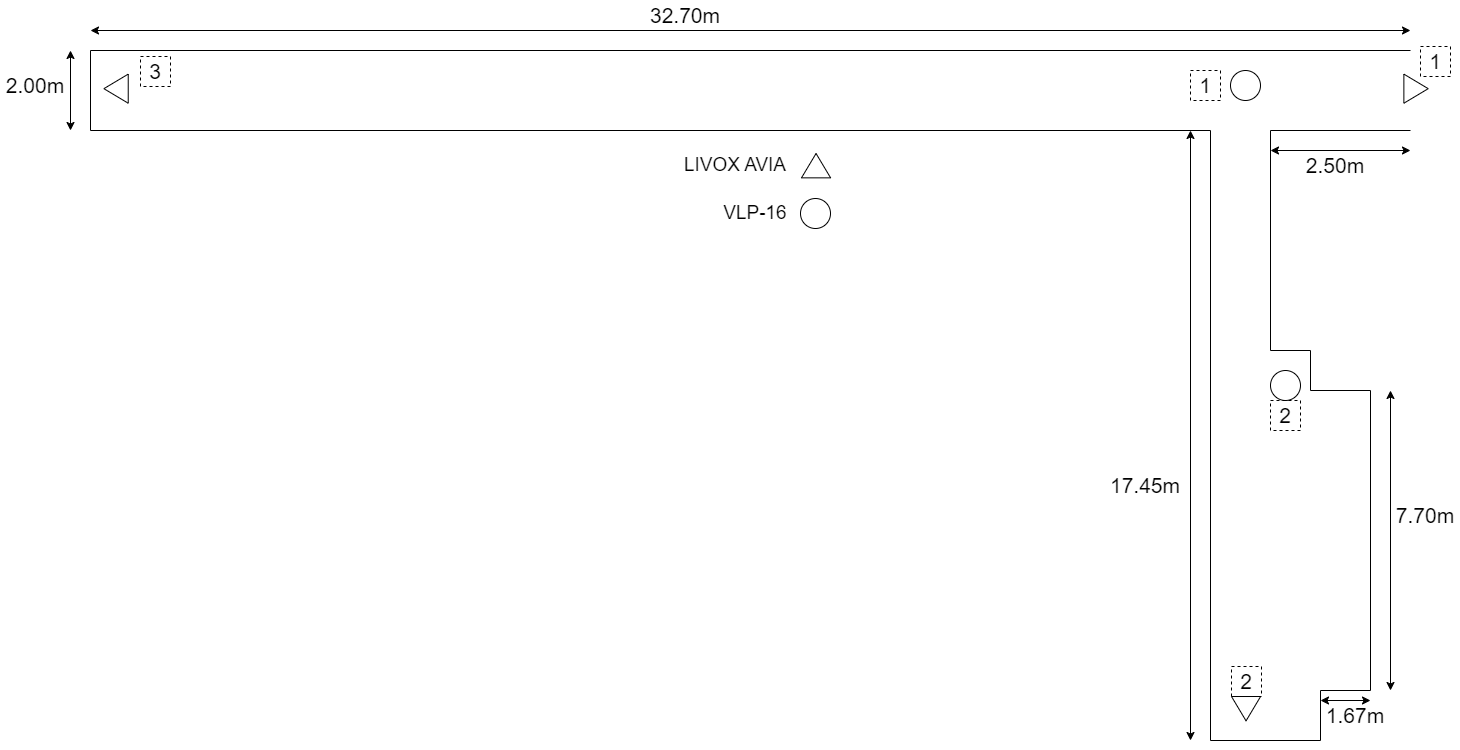}
	\caption{Hallway with position of LiDAR system.}
	\label{fig:lidar:pos:hallway}
\end{figure}

Figure 3 presents a plan‐view layout of a hallway, annotated with three groups of LiDAR sensor stations. The figure illustrates how the LiDAR sensor array is distributed to provide coverage across the entire area. Unlike in the workplace room, this setup includes both Avia and VLP-16 sensors. The Avia sensors are marked with triangles and the VLP-16 sensors with circles, while dashed squares indicate station numbers. Station 1 can be seen in the bottom left part of the figure and station 2 is located in the top part of the hallway. Station 3 is shown at the bottom right part of the figure.

\begin{table}[!htbp]
    \centering
    \caption{Position of the LiDAR sensors in the main room.}
    \begin{tabular}{|l|l|l|l|}
    \hline
        \textbf{LiDAR} &\textbf{ X-coordinate (m)} & \textbf{Y-coordinate (m)} & \textbf{Z-coordinate (m)} \\ \hline
        Avia1 & 1.38 & 7.34 & 0.95 \\ \hline
        Avia2 & 1.70 & 10.33 & 0.97 \\ \hline
        Avia3 & 4.85 & 10.93 & 1.05 \\ \hline
        Avia4 & 8.24 & 8.73 & 1.08 \\ \hline
        Avia5 & 7.73 & 0.14 & 1.57 \\ \hline
        Avia6 & 3.81 & 0.24 & 1.44 \\ \hline

    \end{tabular}
    \label{tab:position:lidar:room}
\end{table}

Table 1 shows the sensor positions for the six Avia LiDAR sensors, installed in the stations~1 to~6 around the room in Figure 2. Each row in the table corresponds to a single sensor, identified by its name in the “LiDAR” column. The X‐ and Y‐coordinates indicate the horizontal placement of the sensor in meters relative to an origin, which is in the bottom left corner, while the Z‐coordinate indicates the sensor’s height above the floor. These coordinates show that sensors are positioned at various points along the perimeter (X and Y-coordinate) and mounted at different heights (Z-coordinate) between roughly 0.9m and 1.8m, mostly on top of the furniture in the area. This distribution ensures coverage across the entire workspace and accommodates potential occlusions from furniture or architectural elements.

\begin{table}[!t]
    \centering
    \caption{Position of the LiDAR sensors in the hallway.}
    \begin{tabular}{|l|l|l|l|}
    \hline
        \textbf{LiDAR} &\textbf{ X-coordinate (m)} & \textbf{Y-coordinate (m)} & \textbf{Z-coordinate (m)} \\ \hline
        Avia1 & 0.00 & 1.00 & 1.30 \\ \hline
        Avia2 & 3.70 & 16.95 & 1.30 \\ \hline
        Avia3 & 32.20 & 1.00 & 1.30 \\ \hline
        VLP1 & 3.70 & 0.50 & 1.30 \\ \hline
        VLP2 & 2.00 & 8.75 & 1.30 \\ \hline
    \end{tabular}
    \label{tab:position:lidar:hallway}
\end{table}

Table 2 shows the sensor positions for the five LiDAR sensors, three Avia units and two VLP units, installed in the stations 1 to 3 around the hallway in Figure 3. Each row in the table corresponds to a single sensor, identified by its name in the “LiDAR” column. Similar to Table 2, the X‐ and Y‐coordinates indicate the horizontal placement of the sensor in meters relative to an origin, which is in the bottom left corner, while the Z‐coordinate indicates the sensor’s height above the floor. LiDAR sensors in the hallway were positioned on the special mounts and are all at the constant height of 1.30 meters.

{In general, sensor mounting height influences the sensor’s field-of-view, incidence angles with surfaces, likelihood of occlusion by furniture or other obstacles, and ultimately the completeness of the point cloud collected. For instance, Garigipati et al.}~\cite{garigipati2022evaluation}{ identify mounting position as a factor for indoor LiDAR algorithm performance, but the influence was only noticed in occluded environments. We applied this reasoning to the furnished main-room environment and by selecting heights between 0.95 m and 1.57 m we reduced occlusion from furniture and improved coverage, whereas in the unobstructed corridor a uniform 1.30 m height was sufficient.}

Figure 4 presents the proposed LiDAR data acquisition and processing pipeline, highlighting both the hardware components and their interactions. In this system, each sensor device integrates a LiDAR unit (either Avia or VLP), which generates raw point clouds that are fed into the edge device in the form of NVIDIA Jetson Nano Developer Kit~\cite{nvidia_jetson_nano}. Within the edge device, a Grabber module continuously retrieves frames from the LiDAR sensor, and a Converter module reformats these frames into a standardized data representation. A Buffer component is then employed to manage high-throughput bursts and mitigate potential network latency by providing temporary storage. Subsequently, a Transformer module applies lightweight processing or compression steps to the data, optimizing it for transmission. The processed LiDAR data are then relayed via the Transmitter module to the edge server through a TP-Link Archer AX73 router (configured as an AP) to facilitate reliable network connectivity. The edge server itself, implemented on an NVIDIA Jetson Xavier NX Developer Kit, receives incoming data through a Receiver module, which hands the data off to an Aggregator responsible for collating, synchronizing, or otherwise fusing data streams from multiple sensor devices. %Finally, all integrated data are written to a hard disk drive (HDD) in the form of a binary file for subsequent analysis.
Finally, all integrated data are stored on a Hard Disk Drive (HDD) in binary format to facilitate subsequent analysis.

\begin{figure}[!b]
	\centering	\includegraphics[height=0.4\linewidth]{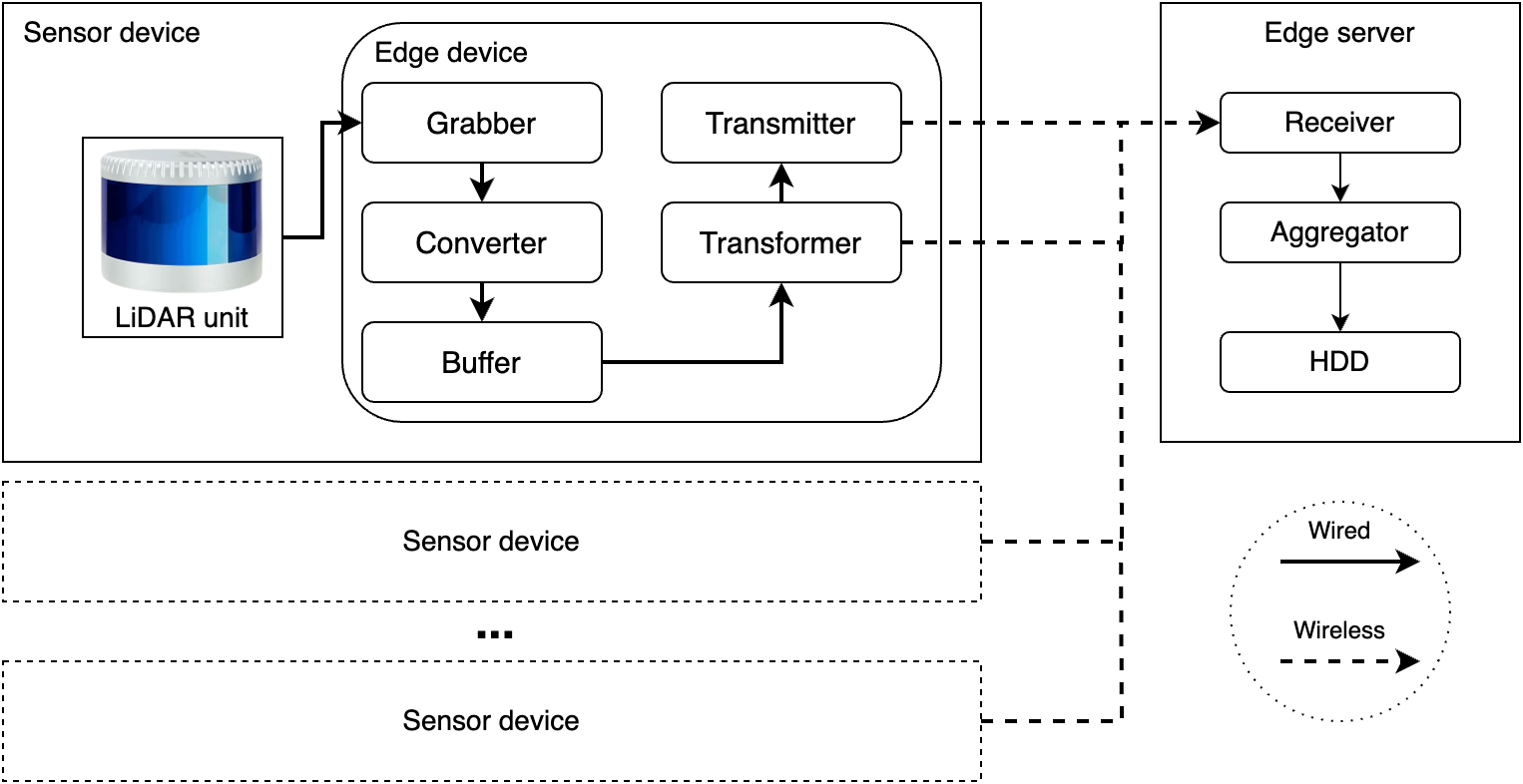}
	\caption{LiDAR sensor data acquisition scheme.}
\label{fig:lidar:archi}
\end{figure}

\subsection*{Measurement System for the RSSI}
\label{sec:RSSI_measurement}
Within the same indoor environment outlined in the previous section, RSSI data is collected at a set of predefined grid locations. The measurement setup employs a {single} TP-Link TL-WR841N device configured as a Wi-Fi AP, operating in the 2.4 GHz frequency band, as shown in Figure 5. This AP is equipped with two omnidirectional antennas, which emit signals uniformly in all directions at the antenna’s elevation.

We collect the RSSI measurements using a commercial Android-based smartphone (Samsung model SM-A556B/DS, running Android 14) with the open-source application Wi-Fi Analyzer~\cite{wifi_analyzer}. The use of widely available, off-the-shelf hardware for both the AP and the measurement device is a deliberate choice aimed at replicating realistic deployment conditions and capturing signal behavior under everyday usage scenarios. {Other wireless networks, WiFi and Bluetooth, are also operating in the same environment, however their influence is considered negligible} \cite{garroppo2011experimental}, {as well as distribution of values of WiFi RSSI in the $2.4GHz$ of around $2dBm$} \cite{park2011implications}, {compared to the differences in spatial changes.}

\begin{figure}[!t]
	\centering	\includegraphics[width=\linewidth]{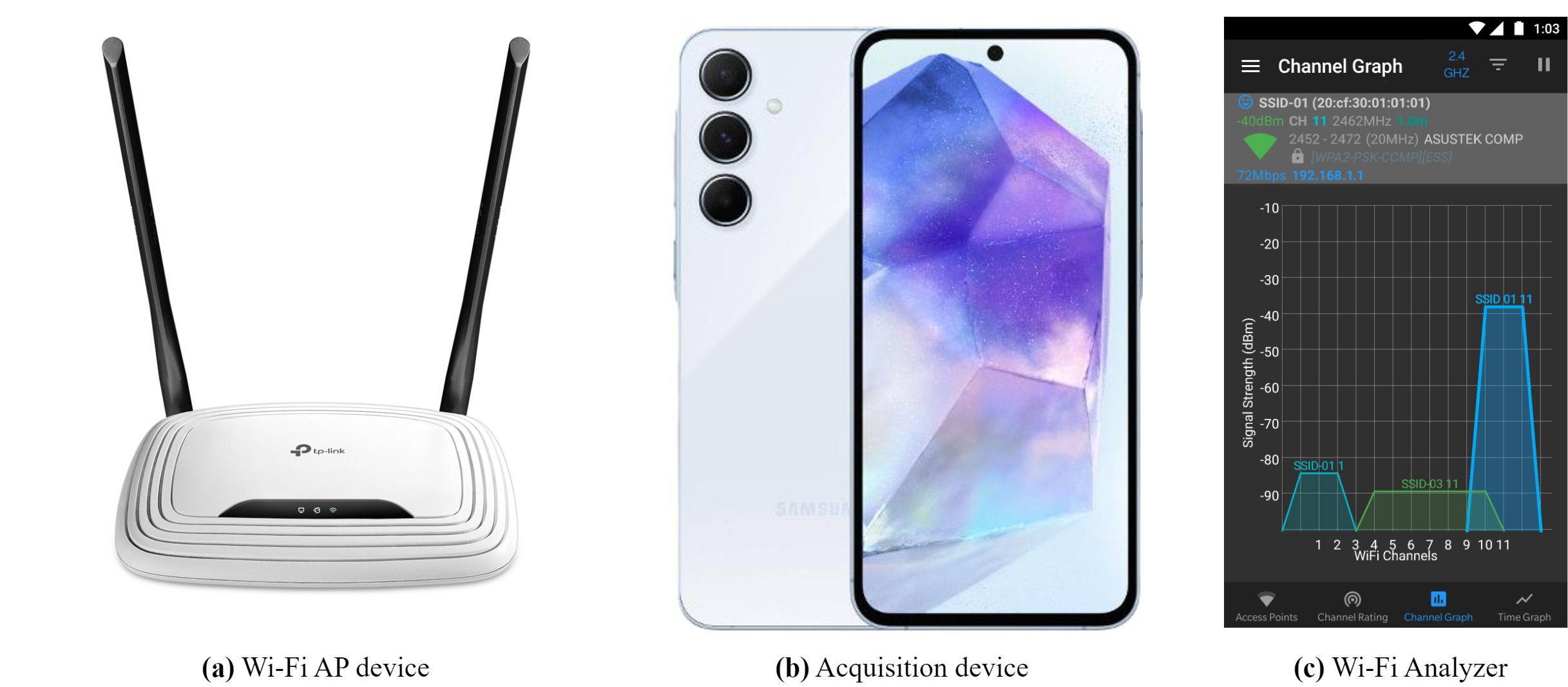}
	\caption{RSSI acquisition hardware and software.}
\label{fig:acquisition_equipment}
\end{figure}

\subsection*{RSSI Measurement Process}

The RSSI measurement process involves positioning a Wi-Fi AP at various locations across a predefined grid and collecting RSSI readings at each corresponding grid point. These readings are gathered using a User Equipment (UE) device in the form of a commercial smartphone held at an operating height of 1.5 meters, reflecting typical user behavior. The AP is placed at a comparable height of 1.2 meters to maintain consistency in signal propagation conditions.

The measurement locations are uniformly distributed throughout the indoor environment, as illustrated in Figure 6, which includes a point cloud representation of the space. An example grid of $53$ measurement points is depicted in Figure 6b, where red dots indicate locations spaced $2$ meters apart in both the $x$ and $y$ directions. {The color variation of the contour plot illustrate the signal strength variation, \textit{blue} marking regions with lower RSSI values, \textit{yellow} marking regions with higher RSSI values.} This spacing was chosen to balance the need for capturing meaningful variations in signal strength while minimizing redundancy and reducing the labor-intensive nature of manual measurements. Due to physical constraints of the office space, the RSSI measurement at location 6 (upper left corner), as an exception to the regular spacing, was taken only 1 meter from location 5, as illustrated in Figure 6b. {The RSSI measurement points deviation is $0.2m$. Empirically it was observed that the RSSI values do not change for such small displacements, with the first observable changes noticed at around 1m displacement. This is due to the limited sensitivity of devices, the software used, and the relatively low variability of the RSSI metric itself} \cite{sadowski2018rssi} {for distances beyond $2m$. The RSSI values depend solely on the relative placement of the AP, the user equipment, and the intervening obstacles (office environment). Consequently, the dataset includes only the RSSI measurements and their associated spatial coordinates, while the temporal aspect is not considered relevant. Thus, in this setup, only spatial synchronization is ensured. The point number $1$ is positioned at a distance of $0.2m$ from both adjacent walls, in the room corner, and the points $2-6$ along the wall at $0.2m$ distance from it, therefore all other measurement points could be referenced to a common coordinate system derived from the 3D data.}

The indoor environment comprises a typical office space (shown in {orange in Figure 6a) and an adjacent corridor and elevator hall} extending southward (depicted in blue). This complex spatial layout serves two purposes: first, it simulates real-world Wi-Fi deployment scenarios; second, it enables the analysis of signal propagation under conditions involving multiple walls and significant spatial separation between the AP and measurement points. Measurement locations were chosen to represent areas of frequent user presence.

\subsection*{Measurement Setup Details}

The measurement sequence is organized numerically, as illustrated in Figure 6b, to enable efficient navigation across the grid and minimize human error during data collection. A total of $20$ distinct setups were created, each comprising $53$ grid locations. While the expected number of RSSI samples is $1,060$ ($20 \times 53$), the actual count is $1,027$ due to weak signals at distant points and occasional human error. Missing values are addressed using preprocessing functions available in a public repository linked in this paper. {In the visualizations, linear interpolation is used to fill the missing values using the \textit{Scipy}} \cite{2020SciPy-NMeth} {library. If needed for specific use case, other interpolation methods are also available for use, namely “nearest”, “slinear”, “cubic”, “quintic” and “pchip”.}

The measurements were conducted under two general scenarios: one with the office completely empty, and another with regular working activity taking place {with 7-10 people present. Each of the 20 setups (12 in the first scenario and 8 in the second)} was designed to reflect realistic signal propagation conditions while maximizing environmental diversity and experimental comparability. {The aim is to provide data that would allow for analysis of how the presence of people during working days affects the signal coverage, compared to an unoccupied office.}

\subsubsection*{Scenario 1: Empty Office (12 Setups)}

In the first scenario, the office space was unoccupied, providing a baseline for signal measurements under static, interference-free conditions. Twelve different measurement setups were defined, each placing the AP at a different location in the room. The chosen positions include locations close to corners (e.g., positions $1$ and $30$), near the walls of the office (e.g., positions $3$, $5$, $18$, $25$), and along the corridor (e.g., positions $33$, $37$, $44$). This arrangement allowed us to capture a wide range of propagation paths, including long-range line-of-sight (in the office and along the corridor), heavily obstructed links (points near the end of the corridor), and reflection-dominated geometries (along the corridor). Special attention was given to placing the AP in setups that reflect practical deployment norms, particularly along room perimeters, without introducing redundancy.

\subsubsection*{Scenario 2: Office with Regular Activity (8 Setups)}

With the second scenario, we aimed to capture the signal strength changes of indoor environments during regular office operations. Here, measurements were collected while 7 to 10 individuals were present in the room, engaged in normal activities such as working at desks, walking, or conversing. This scenario consists of two subgroups. In the first subgroup, four of the AP positions used in Scenario~1 (specifically positions $1$, $18$, $26$, and $30$) were reused under occupied conditions. {The presence of people affects the signal propagation, thus resulting in a new set of (lower dB) values besides the overlapping of the AP and UE locations in the same measurement positions as in Scenario 1. Therefore, the measurements obtained in these four setups are considered unique and new. Such data} enables direct comparisons of signal behavior in identical spatial setups between empty and active environments, isolating the effects of human presence on signal attenuation.

In the second subgroup, four entirely new AP positions were introduced: $20$, $14$, $35$, and $47$. These were selected to represent functionally and structurally distinct zones within the office, such as densely populated desk clusters, regions near shared electronic equipment, and new locations along the corridor. These locations were not used in Scenario 1 and serve to broaden the dataset by introducing previously unobserved signal dynamics. As such, Scenario 2 provides grounds for the exploration and evaluation of potential REM models in new, unseen conditions.

Such a dataset composition ensures coverage of realistic deployment cases, supports comparative analyses, and enhances the representativeness of the measurement campaign for signal modeling and indoor localization tasks.

\begin{figure}[!t]
	\centering	\includegraphics[width=\linewidth]{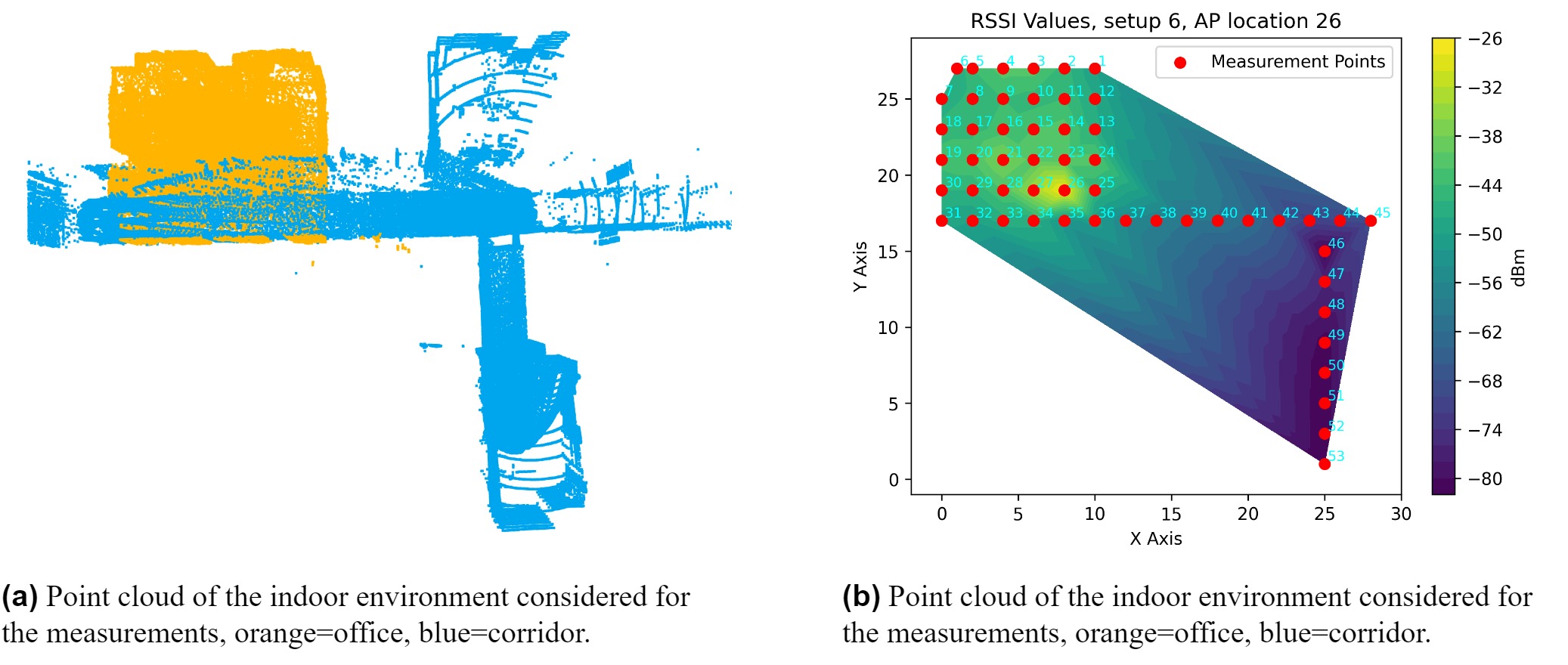}
	\caption{Indoor environment and measurements.}
\label{fig:indoor_measurements}
\end{figure}

\section*{Data Record} % (Data Descriptors only)}
The data generated during the current study is available at \textit{Zenodo} \cite{milosheski2025}.

\paragraph*{3D data}

The point cloud data is provided in three separate files: one containing the office data, another containing the corridor data, and a third \textit{.ply} file representing the combined point cloud. This combined model accurately reflects the physical environment in which the RSSI measurements were collected and is produced through a registration process that aligns the two individual point clouds. Registration is performed manually by selecting corresponding points between the office and corridor datasets, {which in this case are on the common wall that separates them. Example results of the registration process, combining the two separately acquired point clouds, are shown in Figure~6a. Points corresponding to the office are marked in orange, while those representing the corridor and elevator area are marked in blue. The overlapping regions in the figure result from mutual visibility through the glass wall separating the two spaces. Because the LiDAR measures distances to reflective surfaces, it captured corridor points during the office scan and office points during the corridor scan, as the laser beam passed through the glass and reflected off the walls behind it.} As the raw 3D data includes a significant number of outlier points, additional post-processing may be necessary when used for generating precise 3D models of the environment. To support such tasks, we also provide a lightweight toolbox for basic point cloud operations such as loading, visualization, outlier removal, and manual registration, further described in section \textit{Usage notes}.

\paragraph*{RSSI data}

The RSSI data is available in two separate formats. The first one is a \textit{.csv} file (visualized in Table 3) containing the raw measurements of the RSSI, together with other AP-specifics, such as estimated distance to the AP, exact frequency, channel, etc. In another \textit{.h5} file, with contents visualized at the end of this section, the extracted RSSI measurements (\textit{data}) are in matrix format as visualized in Figures 6b and 7, allowing for direct operation on the RSSI data.  Furthermore, labels, regarding the setup number (\textit{setup}), AP location  (\textit{ap-locations}), and measurement flow (\textit{indices}) are also included in the \textit{.h5} file. Regarding the potential reuse, the code for extracting the RSSI data and some simple manipulations, such as visualization of particular setups and filling missing values, will also be provided as part of the toolbox. {Missing data appear mostly on the outermost data acquisition points, such as 50-53 in Figure 6b, due to the excessive distance and multiple walls between the AP and UE.}

\subsubsection*{HDF5 File Structure: \texttt{WiFi\_RSSI\_data2025-05-21\_15-51-16.h5}}
\leavevmode\par\noindent
\begin{forest}
for tree={
    font=\ttfamily,
    grow'=0,
    child anchor=west,
    parent anchor=south,
    anchor=west,
    calign=first,
    inner sep=2pt,
    l sep=10pt,
    edge path={
        \noexpand\path [draw, \forestoption{edge}]
        (!u.parent anchor) -- +(5pt,0) |- (.child anchor)\forestoption{edge label};
    },
    before typesetting nodes={
        if n=1
            {insert before={[,phantom]}}
            {}
    },
    fit=band,
}
[/
  [ap\_locations, tier=dataset]
  [data, tier=dataset]
  [indices, tier=dataset]
  [setup, tier=dataset]
]
\end{forest}

\begin{table}[!t]
\centering
\caption{Visualization of a truncated sample of the .csv file.}
\begin{tabular}{cllcll}
\toprule
\# & Time Stamp & SSID & \ldots & Fast Roaming & Setup \\
\midrule
0 & 2024/10/21-11:21:51 & TP-Link\_4462 & \ldots & \textbackslash n & 1 \\
1 & 2024/10/19-10:14:38 & TP-Link\_4462 & \ldots & \textbackslash n & 1 \\
2 & 2024/10/19-10:15:45 & TP-Link\_4462 & \ldots & \textbackslash n & 1 \\
3 & 2024/10/19-10:16:09 & TP-Link\_4462 & \ldots & \textbackslash n & 1 \\
4 & 2024/10/19-10:16:41 & TP-Link\_4462 & \ldots & \textbackslash n & 1 \\
\multicolumn{6}{c}{\textellipsis} \\
\bottomrule
\end{tabular}
\label{tab:csv_sample}
\end{table}

\section*{Technical Validation} % (Data Descriptors only)}

Figures 7a and 7b illustrate the RSSI distributions for measurements taken with the AP placed at location~1, both in the presence and absence of people in the office. In the first scenario (Figure 7a), which corresponds to an unoccupied environment {(no people)}, the signal strength appears noticeably higher toward the center of the office. In contrast, Figure 7b shows reduced signal strength in the same region when people are present. This attenuation is expected, as human bodies absorb and scatter radio signals, leading to reduced RSSI values. 

Another notable phenomenon is the observation of relatively weaker or comparable RSSI values at the AP’s exact location, compared to immediately adjacent positions. For example, in setup 2 (Figure 7c), the AP is placed at location 3, where an RSSI of $-33dBm$ is recorded. Interestingly, the neighboring location 2 shows a significantly stronger signal of $-19dBm$. This effect may be attributed to the vertical proximity of the measurement device to the AP, capturing a point within the antenna’s radiation null. Omnidirectional antennas typically exhibit a {"donut-shaped"} radiation pattern, which results in weaker signal strength directly above or below the AP.

A similar pattern is evident in setup 19 (Figure 7c), where the AP is placed at location 35 and a comparable RSSI value of $-29dBm$ is measured at adjacent location 36. This can likely be explained by the confined geometry of the corridor, where multipath propagation and surface reflections contribute to localized signal enhancements.

Setups 1 and 13 share the same AP location but differ in human occupancy. Notably, both setups record identical RSSI values at position 1, which coincides with the AP’s exact location. While this might initially suggest a lack of environmental impact, it is consistent with expectations. Since the AP resides in an open environment and position 1 corresponds to its exact placement, the presence of people is unlikely to substantially influence the signal at that point due to the minimal spatial separation between transmitter and receiver.

{Given the dataset’s spatial consistency between the RSSI measurements and 3D representation of the environment, it is well suited for training and validating REM estimation models} \cite{cisse2023irgan, bakirtzis2022deepray}. {Unlike previous works using 2D input images with AP locations, this dataset provides 3D input data that captures detailed environmental structure along with actual RSSI measurements serving as ground truth for REM estimation. While simulated data can support large-scale initial training} \cite{bakirtzis2024indoor}, {this dataset enables realistic validation. Measurements include both human-present and human-free conditions, allowing evaluation of model robustness. The dataset is organized for progressive validation, from pre-deployment testing (setups 1–12) to deployment fine-tuning (setups 13–16) and evaluation in unseen deployment scenarios (setups 17–20).}

\begin{figure}[!t]
	\centering	\includegraphics[width=\linewidth]{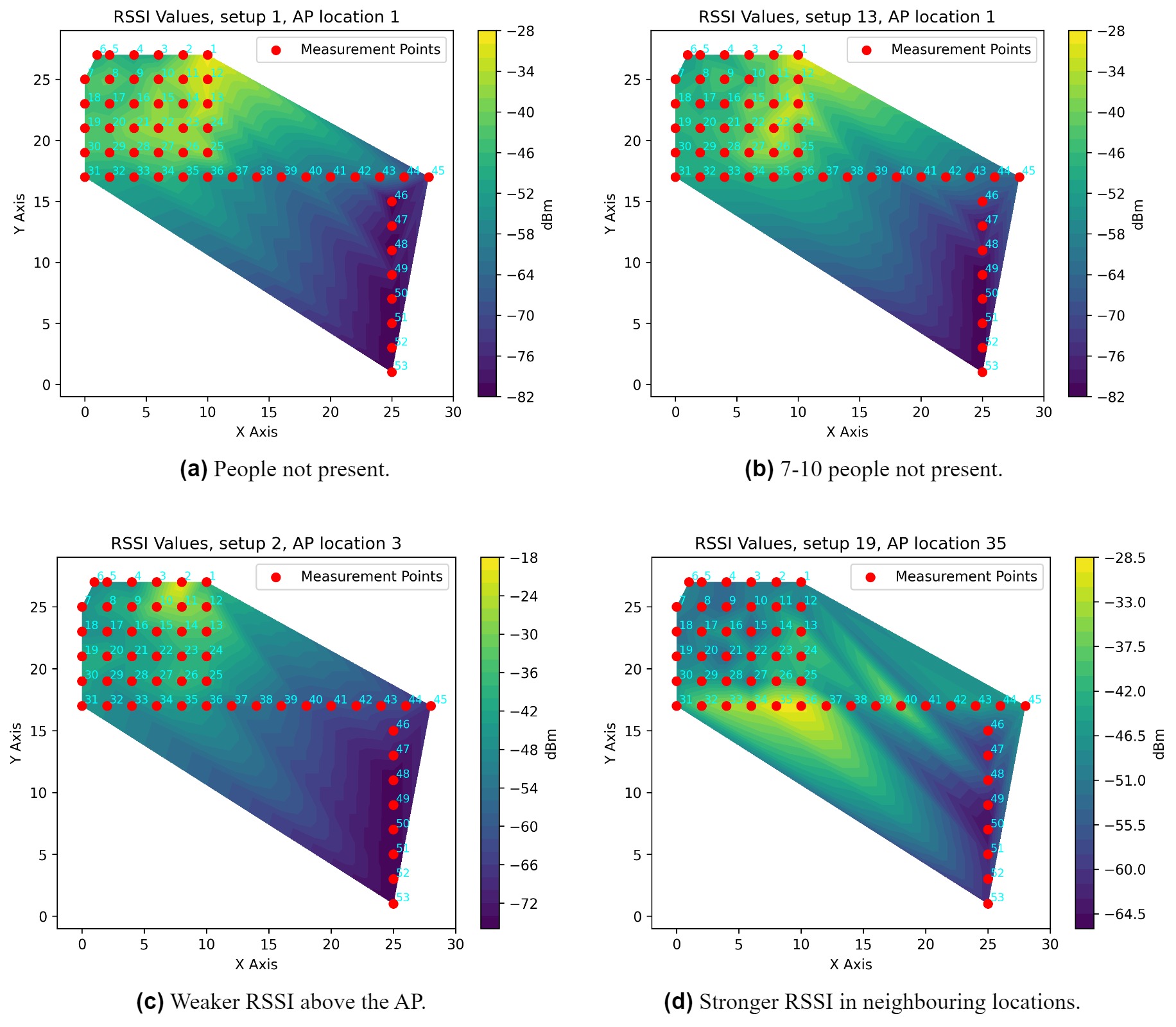}
	\caption{Observations of the contour plots of RSSI measurements.}
\label{fig:indoor_measurements}
\end{figure}

\section*{Usage Notes} 

The intended usage of the dataset is to aid the development of advanced ML models, such as the ones based on GAI, for accurate REM 
estimation based on 3D environmental data, which can correlate the spatial signal measurements with the physical structure of the environment. This is particularly important for supporting applications that demand consistently high throughput and ultra-low latency, such as the XR-related devices. {Furthermore, it also allows for analysis of how the presence of people in the office during working days affects the signal coverage. Although the number of people present and their uncontrolled activities, such as walking and sitting, influence signal propagation, these factors were not controlled in the current study and will be systematically addressed in future measurement campaigns.}

{The dataset files, including the RSSI measurements and the 3D point cloud, together with the software for data manipulations, which will be detailed in the next section, are contained in a single \textit{.zip} directory. After downloading, users can open the project in an interactive development environment (IDE) such as \textit{PyCharm} and install the required libraries listed in the \textit{requirements.txt} file. Once installed, the \textit{example\textunderscore operations.ipynb} notebook provides sample workflows for loading and processing both data types.}

We provide the point cloud data in two separate files: one representing the office space and the other the corridor. Because the RSSI measurements are environment-specific, the dataset supports various usage configurations. In simpler scenarios, researchers can use only the office data to study signal propagation in open indoor areas, both with and without human presence. Alternatively, the corridor data can be used independently to analyze long line-of-sight signal behavior and reflection-rich conditions, due to the corridor’s narrow width of approximately $2m$. More complex experimental setups can utilize the full dataset by combining both the office and corridor data, including cases where multiple walls separate the AP and UE, enabling the study of challenging multipath and obstruction scenarios.

To support the manual inspection and alignment of 3D point cloud data, we developed a lightweight interactive toolbox built on the Open3D~\cite{zhou2018open3d} library. It enables users to visually crop regions of interest within point clouds, select key correspondence points across multiple scans, and perform rigid registration based on these manual inputs. The tool also allows for real-time visualization of registration results and supports the export of aligned point cloud data for further use.

Furthermore, {the toolbox also provides} a set of functions for basic Wi-Fi data processing, such as loading, organization, and visualization, which help analyze the signal propagation in the indoor environment. The RSSI measurements are loaded from structured log files, which are collected across multiple experimental setups and parsed into a unified tabular format. Signal strength values are extracted and spatially mapped onto a predefined two-dimensional grid that reflects the physical layout of the environment, including a single office room and two corridors. The toolbox accounts for missing or corrupted entries through interpolation to maintain data consistency. For visualization, it generates contour plots that illustrate spatial signal distribution and measurement density, along with annotations indicating measurement indices and AP locations. {Contour plots are generated using the \textit{matplotlib}}\cite{Hunter:2007} {library, using the \textit{tricontourf} function, based on linear interpolation, and are only for visualization purposes. Other methods could also work.} To enable reproducibility and efficient storage, the processed datasets, including signal maps, setup identifiers, and spatial indices, are compiled into compressed HDF5 \cite{hdf5} files with timestamped filenames.

\section*{Data availability}
The data produced during this study is publicly available on \textit{Zenodo} \url{https://doi.org/10.5281/zenodo.15791300}, in the \textit{data} folder. It consists of the 3D point clouds in three separate \textit{ply} files, which include the measurement area, consisting of two point clouds, and one example output of the point cloud registration, produced by the \textit{example\_operations.ipynb} script. The RSSI data is provided in a concise \textit{csv} file and in raw \textit{h5} format. The repository organization is shown in Figure 8.

\begin{figure}[!ht]
        \centering
	\includegraphics[width=0.6\linewidth]{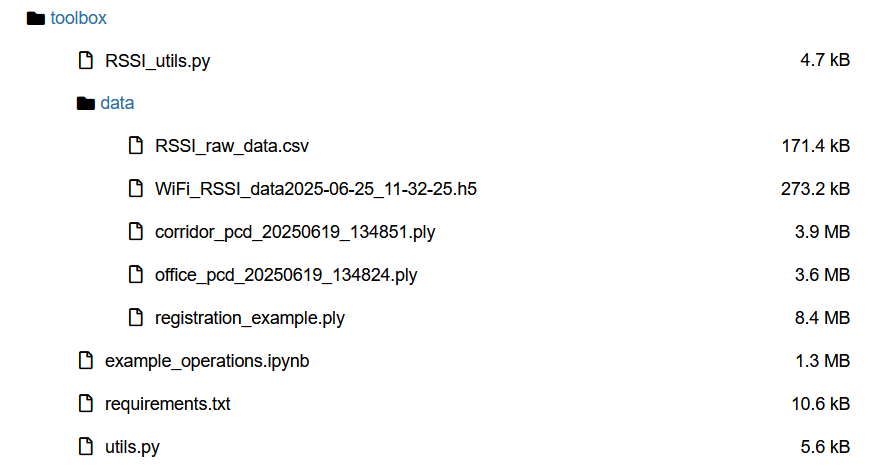}
	\caption{Organization of the dataset repository.}
	\label{fig:repo_organization}
\end{figure}

\section*{Code availability}
The code supporting the findings of this study is publicly available on \textit{Zenodo} \url{https://doi.org/10.5281/zenodo.15791300}. It includes scripts for loading, visualizing, and processing the 3D point cloud and RSSI data.

\section*{Acknowledgments}% (not compulsory)}

This work was supported in part by the Slovenian Research Agency (ARIS) under grants P2-0016 and MN-0009-106 and by the Japan Society for the Promotion of Science (JSPS) under grants 23H00464, 25H01124, and 120245002.
Additional support was provided by the bilateral project MISA (BI-JP/24-26-001), funded jointly by ARIS and the JSPS.

\section*{Author contributions statement}

L.M. conceived the experiment,  L.M., K.A. and B.B. conducted the experiment and analysed the results. J. H., and R. S. supervised the experimental progress and provided project support. All authors reviewed the manuscript.

\section*{Competing Interests}

The authors declare no competing interests.

\end{document}